\documentclass[dvips,12pt]{article}     
\usepackage{color}
\usepackage{hyperref}
\usepackage{amsfonts}
\usepackage{amsmath, amssymb}
\usepackage{graphicx}
\usepackage{psfrag}
\usepackage{youngtab}
\Yboxdim{8pt}  

\newcommand{\ra}{\rightarrow}

\newcommand{\be}{\begin{equation}}
\newcommand{\ee}{\end{equation}}
\newcommand{\ba}{\begin{eqnarray}} 
\newcommand{\ea}{\end{eqnarray}}
\newcommand{\bi}{\begin{itemize}}
\newcommand{\ei}{\end{itemize}}

\newcommand{\Tr}{{\rm Tr}}

\newcommand{\R}{{\rm R}}

\newcommand{\Ncal}{{\mathcal N}}

\newcommand{\Ocal}{{\mathcal O}}

\newcommand{\nn}{\nonumber}
\newcommand{\mo}{{-1}} 

\newcommand{\f}{\frac}

\newcommand{\oo}{\frac{1}}

\def\Dslash{\,\,{\raise.15ex\hbox{/}\mkern-12mu D}}
\def\Dbarslash{\,\,{\raise.15ex\hbox{/}\mkern-12mu {\bar D}}}
\def\delslash{\,\,{\raise.15ex\hbox{/}\mkern-9mu \partial}}
\def\delbarslash{\,\,{\raise.15ex\hbox{/}\mkern-9mu {\bar\partial}}}
\def\pslash{\,\,{\raise.15ex\hbox{/}\mkern-9mu p}}
\def\calDslash{\,\,{\raise.15ex\hbox{/}\mkern-12mu {\cal D}}}

\renewcommand{\bar}{\overline}

\begin{document}
\baselineskip=15.5pt
\renewcommand{\theequation}{\arabic{section}.\arabic{equation}}
\pagestyle{plain}
\setcounter{page}{1}
\bibliographystyle{utcaps}
\begin{titlepage}

\rightline{\small{\tt NSF-KITP-07-164}}
\begin{center}

\vskip 3 cm

{\Large {\bf  A prediction for bubbling geometries}}

\vskip 4cm
Takuya Okuda

\vskip 1cm

Kavli Institute for Theoretical Physics

University of California

Santa Barbara, CA 93106, USA

\vskip 4cm

{\bf Abstract}

\end{center}

We study the supersymmetric
circular Wilson loops in  $\Ncal=4$ Yang-Mills theory.
Their vacuum expectation values are computed
in the parameter region that admits smooth bubbling geometry duals.
The results are a prediction for the supergravity action evaluated
on the bubbling geometries for Wilson loops.

\end{titlepage}

\newpage


\tableofcontents
\section{Introduction and summary}

In a holographic correspondence,  a quantum gravitational
system is exactly equivalent to a non-gravitational theory
on a lower dimensional space.
In known examples \cite{Maldacena:1997re,Gopakumar:1998ki}
the non-gravitational systems are gauge theories.
Gauge invariant observables are mapped
to non-normalizable deformations of the gravitational background. 

\begin{figure}[ht]
\centering
\psfrag{a}{
\begin{tabular}{c}
Fundamental\\
string
\end{tabular}}
\psfrag{b}{
\begin{tabular}{c}
D3-brane on \\
$S^2\subset AdS_5$
\end{tabular}}
\psfrag{c}{
\begin{tabular}{c}
D5-brane on\\
$S^4\subset S^5$
\end{tabular} }
\psfrag{d}{\hspace{1.5mm}Bubbling geometry}
\includegraphics[scale=.45]{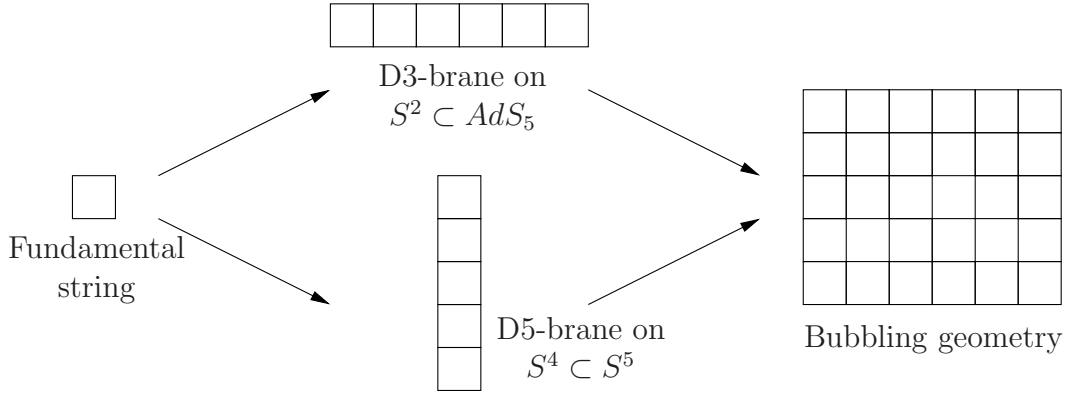}
\caption{Gravity duals of Wilson loops in $\Ncal=4$ Yang-Mills.
A string corresponds to the fundamental representation,
a D3-brane to a symmetric representation,
a D5-brane to an anti-symmetric representation,
and a bubbling geometry to the representation specified 
by a rectangular Young diagram.
(D3-brane ..., D5-brane ...) in the figure should be replaced by (D3-brane wrapping $S^3\subset AdS_5$, D3-brane 
wrapping $S^3\subset S^5$)
for local operators in $\Ncal=4$ Yang-Mills, and by (D-brane, anti-D-brane)
for Wilson loops in Chern-Simons theory.
}
\label{young-trans}
\end{figure}

Following progress in understanding the gravity duals of
local operators in a gauge theory, the recent years have seen further
extension of the dictionary to the realm of non-local gauge invariant operators.
The picture that has emerged is strikingly universal for all local and non-local
operators.
Consider operators of the form $\Tr_R(...)$, where the trace is
evaluated in a representation $R$ 
of $SU(N)$.
We will often use the symbol $R$ to denote the associated
Young diagram as well.
For different representations $R$, the operators are best described by
different objects on the gravity side.
The fundamental representation $R=\yng(1)$
corresponds to a fundamental string.
A high-rank symmetric representation corresponds to a D-brane, while an 
anti-symmetric representation is given by another type of D-brane.
Finally, an operator with a large rectangular Young diagram is dual
to a smooth bubbling geometry with a flux supported on a new cycle.
A general Young diagram corresponds to a combination of these objects.
This pattern, summarized in Figure \ref{young-trans}, has been demonstrated for
local operators \cite{McGreevy:2000cw,Hashimoto:2000zp,
Balasubramanian:2001nh,Corley:2001zk,Lin:2004nb} 
and Wilson loops \cite{Maldacena:1998im,Rey:1998ik,Drukker:2005kx,
Gomis:2006sb,
Yamaguchi:2006tq,
Hartnoll:2006is,Yamaguchi:2006te,
Gomis:2006im,
Lunin:2006xr,D'Hoker:2007fq} in $\Ncal=4$ Yang-Mills, 
as well as for Wilson loops in Chern-Simons theory
\cite{Gopakumar:1998ki, Gomis:2006mv,Gomis:2007kz,Okuda:2007ai}.
Higher dimensional non-local  operators \cite{Gukov:2006jk,Gomis:2007fi,
Karch:2000gx,Lunin:2006xr,Gomis:2006cu,Lunin:2007ab} 
in gauge theories
have descriptions in terms of branes and bubbling geometries.

The identification of the operators and their dual bubbling geometries
has been made on the basis of symmetries and charges.
An interesting test of the identification, and AdS/CFT itself,
would be to compare the expectation value of an operator
with the on-shell supergravity action evaluated on the bubbling geometry.
For most operators, however, the test is trivial.
the expectation value vanishes for local operators, 
and does not depend on the coupling for the straight Wilson line.
Even if the expectation value does not vanish, the computation
in the gauge theory is usually done only in weak coupling.
It is then hard to make comparison with the strong coupling
computation performed on the gravity side.

\begin{figure}[htbb]
\begin{center}
\psfrag{n1}{$n_1$}
\psfrag{k1}{$k_1$}
\psfrag{n2}{$n_2$}
\psfrag{kg-1}{$k_{g-1}$}
\psfrag{ng}{$n_g$}
\psfrag{kg}{$k_g$}
\psfrag{ng+1}{$n_{g+1}$}
\includegraphics[width=80mm]{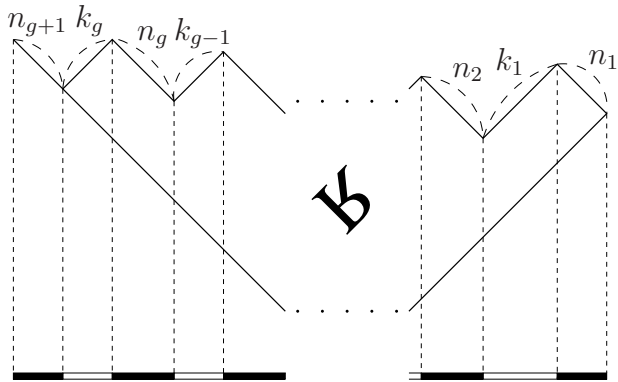} 
\caption{The Young diagram $R$, shown rotated and inverted, 
is specified by the lengths $n_I$ and $k_I$ of the
edges.
Equivalently, $n_I$ and $k_I$ denote the lengths 
of the black and white regions in
the Maya diagram.
$n_{g+1}$ is defined by $\sum_{I=1}^{g+1} n_I=N$.
}
\label{parametrization}
\end{center}
\end{figure}

 In this paper we make a non-trivial prediction for the bubbling geometries
dual to particular Wilson loops  in $\Ncal=4$ Yang-Mills,
by computing  the vevs of the operators on the gauge theory side
{\it in the strong coupling regime}.
The operators are the circular supersymmetric Wilson loops
defined as
\ba
W_R\equiv
\Tr_R P \exp \oint (A+\theta^i \Phi^i ds). \label{Wilson-loop}
\ea
Here $A=A_\mu dx^\mu$ is the gauge field,
$\Phi^i$ ($i=1,...,6$) are the real scalars,
$s$ is a parameter for a circle in Euclidean $\R^4$ such that
$||dx/ds||=1$,
and $(\theta^i)$ is a constant unit vector in $\R^6$.
A smooth bubbling geometry is expected to be dual to a circular Wilson loop
whose Young diagram has long edges.
More precisely, if we parametrize the diagram as 
in Figure \ref{parametrization},
$n_I$ for $I=1,...,g+1$ and $k_I$ for $I=1,...,g$ all  have to be of order $N$ for the
geometry to be smooth.  
The circular Wilson loop is very special
in that such strong coupling computation from gauge theory is possible,
due to the conjecture \cite{Drukker:2000rr} that 
the Gaussian matrix model captures the circular Wilson loops
to all orders in $1/N$ and the 't Hooft coupling $\lambda=g_{YM}^2 N$.

Relegating the computations to later sections,
let us simply summarize the main result:
in the limit $N\ra\infty$ with $n_I/N$ and $k_I/N$ kept finite,
and for the large finite 't Hooft coupling $\lambda$,
the vev of the circular Wilson loop in the Yang-Mills
with gauge group $SU(N)$ is given by 
\ba
\langle W_R\rangle
&=&
\exp\left( \f{\lambda}{8N} \sum_{I=1}^{g+1} n_I (K_I-|R|/N)^2 
+\Ocal(N^2\log \lambda)\right). 
\label{the-results}
\ea
Here $|R|$ is the n-ality, i.e.,  the number of boxes in $R$.
We have also defined $K_{I}=k_{I}+k_{I+1}+...+k_{g}$
for $1\leq I\leq g$ and $K_{g+1}=0$.
An important check of the result is 
that it is invariant under 
complex conjugation $R\ra \bar R$ of the representation, 
or in terms of the parameters, under
\ba
n_I \ra  n_{g+2-I},~~
k_{I}\ra k_{g+1-I}.
\ea
We expect that the Wilson loop vev is related to the on-shell supergravity
action $S_{\rm sugra}$ as
\ba
\langle W_R\rangle \sim e^{-S_{\rm sugra}}.
\ea
Note that $\langle W_R\rangle =e^{\Ocal ( N^2 \lambda)}$ in our regime.
The quadratic dependence on $N$ of the exponent is
what is expected from the supergravity action
$S_{\rm sugra}=(1/2\kappa^2)\int \sqrt{g}R+...$
with $\kappa^2\sim g_s^2\alpha'^4$, to be evaluated
on the dual geometry.

This result is obtained
by studying the Gaussian matrix model 
that captures the correlation functions
of circular Wilson loops.
In fact we turn the Gaussian matrix model
with operator insertions into
multi-matrix models whose partition functions
are the Wilson loop vevs.
We perform the saddle point analysis
and propose the eigenvalue distributions.
Based on this proposal, we are able to calculate
the Wilson loop vevs with the result (\ref{the-results}).

The circular Wilson loops in Euclidean $\R^4$
can also be thought of as operators in
the  $\Ncal=4$ Yang-Mills defined on $S^4$ \cite{Drukker:2000rr}.
The loop is now the equator of $S^4$, and
the $SO(2)\times SO(3)$ subgroup 
of the $SO(5)$ isometry group is preserved.
The $SO(2)$ is part of the preserved
subgroup $SU(1,1)$ of the conformal group $SO(1,5)$.
The operator (\ref{Wilson-loop}) also preserves
an $SO(5)$ subgroup of the $SO(6)$ R-symmetry group.
Thus the bubbling geometry duals of the Wilson loops
must solve the BPS equations of the type IIB supergravity
with the $EAdS_2 \times S^2 \times  S^4$ ansatz.
Note that we need to work with Euclidean signatures
since the matrix model captures the circular loop
in Euclidean $\R^4$ only.
The resulting geometries should be the Euclidean continuation of the
solutions recently found in \cite{D'Hoker:2007fq}.

Given the result of the present paper,
the challenge is to use the dual geometries to reproduce the
prediction (\ref{the-results}) by computing the on-shell 
supergravity action \cite{bub}.
One would need to deal with such subtle issues as
volume regularization, counter terms, and ambiguity of the action
due to self-duality of the 5-form flux.\footnote{
The parallel problem of matching the Wilson loop vev and the
closed string partition function has been solved 
for the Chern-Simons/conifold duality to all orders
in genus expansion \cite{Gomis:2006mv}.
The matching of the Wilson loop vevs and the dual on-shell
D-brane actions
has also been demonstrated \cite{Drukker:2005kx,Yamaguchi:2006tq,
Hartnoll:2006is} for $\Ncal=4$ Yang-Mills.
}

Though the analysis of 
the matrix models for  $\Ncal=4$ Yang-Mills
is in principle sufficient,
much confidence in the results and the proposed eigenvalue distributions
comes from the study of surprisingly analogous
matrix models that describe the Wilson loops 
in Chern-Simons theory \cite{Marino:2002fk,Aganagic:2002wv,cs-mat}. 
For these matrix models, the resolvents and the spectral curves
can be computed exactly in the limit $N\ra\infty$,
with $n_I/N$ and $k_I/N$ kept finite, 
without the assumption that the 't Hooft coupling is large.
The eigenvalue distributions turn out to be 
qualitatively the same for the $\Ncal=4$
Yang-Mills and Chern-Simons theory.
 
The paper is organized as follows.
In section \ref{matrix-wilson}, we review
the correspondence between the circular
Wilson loops in $\Ncal=4$ Yang-Mills and
the observables in the Gaussian matrix model.
In section \ref{rectangle} we analyze
the Wilson loop with a rectangular
Young diagram.
Section \ref{general} deals with the Wilson loops
in  general representations that admit 
smooth dual bubbling geometries.

\section{Gaussian matrix model  for circular Wilson loops}
\label{matrix-wilson}

It is believed \cite{Drukker:2000rr,Erickson:2000af}
that the correlation functions of  circular
loops in $\Ncal=4$ Yang-Mills are
captured by the Gaussian matrix model.
The precise correspondence states 
in particular that
\ba
\left\langle \Tr_R P \exp \oint (A+\theta^i X^i ds)\right\rangle_{U(N)}
=
\oo Z \int dM \exp\left({-\f{2N}\lambda \Tr M^2}\right)
\Tr_R e^M. \label{mat-wil-U}
\ea
The left-hand side is the normalized expectation value
in the Yang-Mills with gauge group $U(N)$.
The right-hand side is normalized by using
the partition function $Z$ which is
the integral without the insertion of $\Tr_R e^M$.
$dM$ is the standard hermitian matrix measure.
In the absence of operator insertions, the eigenvalues
are distributed according to the Wigner semi-circle law
in the large $N$ limit.
The eigenvalue density $\rho(m)$, normalized so that $\int \rho(m) dm=1$,
is given by
\ba
\rho(m)=\f{2}{\pi\lambda}\sqrt{\lambda-m^2}.
\ea

We are also interested in the $SU(N)$ Yang-Mills theory,
for which the conjecture in \cite{Drukker:2000rr} states that
\ba
\left\langle \Tr_R P \exp \oint (A+\theta^i X^i ds)\right\rangle_{SU(N)}
=
\oo Z \int dM \exp
\left(-\f{2N}\lambda \Tr M^2\right)
\Tr_R \hspace{.4mm} e^{M'}. \label{mat-wil-SU}
\ea
Note that on the right-hand side we need the traceless part
\ba
M':=M-\oo{N}\Tr M\cdot 1_N \label{Mprime}
\ea
of the matrix $M$ so that $e^{M'}\in SU(N)$.
The relation
\ba
\Tr_R\hspace{.4mm} e^{M'}= (\det e^M)^{-|R|/N} \Tr_R e^M  \label{TrM'}
\ea
will be useful later.

Matrix integrals 
have been studied intensively in the past,
partly motivated by non-critical string theory
\cite{Ginsparg:1993is, Francesco:1993nw}
and relation to supersymmetric gauge theories
\cite{Dijkgraaf:2002fc,Dijkgraaf:2002dh}.
In these contexts one does not encounter operators of the form
$\Tr_R e^M$ with the Young diagram $R$ that is large
both in the row and column directions,
and it appears that there is no study of them in the literature.\footnote{
I thank E. Martinec for a discussion on this point.}

\section{Rectangular Young diagram}\label{rectangle}

\begin{figure}[htbb]
\begin{center}
\psfrag{l}{$k$}
\psfrag{n}{$n$}
\includegraphics[width=50mm]{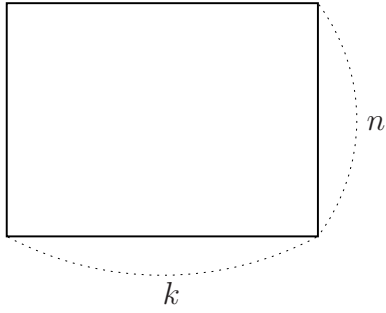}  
\caption{A rectangular Young diagram with $n$ rows and $k$ columns.
}
\label{rectan}
\end{center}
\end{figure}

In this section, we formulate
2-matrix models whose partition functions
are the vev of the Wilson loop
with a rectangular Young diagram shown in Figure \ref{rectan}.
We will have two such matrix models.
The first captures the geometric transition
of D5-branes wrapping $S^4\subset S^5$ in $AdS_5\times S^5$.
The second is similarly interpreted in terms of
D3-branes wrapping $S^2\subset AdS_5$.
The techniques we introduce here are a generalization
of the methods used in \cite{Hartnoll:2006is}.
The pictures that emerge are similar to those of
\cite{Yamaguchi:2006tq,Yamaguchi:2007ps}.

\subsection{Eigenvalues as D5-branes} \label{rec-D5}
We begin with the Yang-Mills with gauge group $U(N)$.
Let $U$ be a unitary matrix 
and consider a sum over all the Young diagrams for which 
the summand is non-vanishing:
\ba
\sum_R
 \Tr_R e^M
\Tr_{R^T} U 
=
\det(1+
e^M\otimes U).\label{RRT-det}
\ea
Here $R^T$ is the transpose of $R$.
This identity, expressed in terms of Schur polynomials
as 
\ba\sum_R s_R(x)s_{R^T}(y)=\prod_{i,j}(1+x_iy_j)
~~\hbox{ for } x=(x_i) \hbox{ and } y=(y_j),
\ea
is well-known \cite{Macdonald}.    
The relation can be inverted:
\ba
\Tr_R e^M=\int dU \det(1+e^M\otimes U^\mo)\Tr_{R^T}U. \label{inversion}
\ea
Here $dU$ is the Haar measure on the unitary group normalized
so that $\int dU=1$.
We now specialize to the rectangular Young diagram
$R$ with $n$ rows and $k$ columns.
We choose $U$ to be a $k\times k$ unitary matrix.
This choice has a considerable advantage.
Let us recall the character formula \cite{Fulton}
\ba
\Tr_R X=\f{\det(x_b{}^{n-a+R_a})}{\det(x_b{}^{n-a})}~~
\hbox{for $X={\rm diag}(x_1,...,x_k)$ and arbitrary $R$},\label{character}
\ea
where $R_a$ is the number of boxes in the $a$-th row of $R$.
It implies that 
\ba
\Tr_{R^T} U=(\det U)^n \label{tr-det}
\ea
for the rectangular diagram $R$ and {\it for
the $k\times k$ matrix} $U$.\footnote{
This can also be understood as follows.
A column of length $k$ gives the rank-$k$ anti-symmetric
representation $A_k$.
The rectangular diagram $R$ is obtained
by symmetrizing $n$ copies of $A_k$.
The representation $A_k$ for $U(k)$ is one dimensional
and the trace in $A_k$ is the determinant.
Thus $\Tr_{R^T} U=\Tr_{{\rm Sym}^n (A_k)} U
=\Tr_{A_k{}^{\otimes n}}U=(\det U)^n$.
}
By substituting (\ref{inversion}) and (\ref{tr-det})
into (\ref{mat-wil-U}), we conclude that
\ba
\langle W_R\rangle_{U(N)}&=&\oo Z \int  dM dU  \exp\left(
-\f{2N}{\lambda}\Tr M^2\right)
  \det(1+ e^M\otimes U^\mo)
(\det U)^n.
\ea
By diagonalizing the matrices as
\ba
M={\rm diag} (m_i)_{i=1}^N,~U={\rm diag} (e^{u_a})_{a=1}^k,
\ea
and redefining $U\ra - U$,
the vev can be written as
\ba
\langle W_R\rangle_{U(N)}&\propto&
\int \prod_a du_a\prod_i dm_i
 \exp\Bigg[ -\f{2N}{\lambda}\sum_{i=1}^N
 m_i^2+\sum_{i<j}\log(m_i-m_j)^2+n\sum_{a=1}^k u_a\nn\\
 &&~~~+ \sum_{a<b}\log\left( 2\sinh\f{u_a-u_b}{2}\right)^2
 +\sum_{a,i} \log(1- e^{m_i-u_a} )\Bigg].
 \label{vev1} 
\ea
The constant of proportionality is independent of $k$ and $n$,
is negligible in the precision we work with, and is thus dropped.
Since the integrand is analytic, the contours of integration
may be deformed.
Each $u_a$ is integrated over a period $2\pi i$.

The leading behavior of the Wilson loop vev in the large $N$ limit can be 
computed in the saddle point approximation.
Note that all the terms in the exponent in (\ref{vev1}) are of order $N^2$.
We thus expect a back-reaction of the $m$-eigenvalue
distributions to the $u$-eigenvalues.
This is in contrast to the case with a single row or column
\cite{Hartnoll:2006is}.
The saddle point equations are
\ba
-\f{4N}\lambda m_i+2\sum_{j\neq i}\oo{m_i-m_j}-\sum_a\f{1}{e^{u_a-m_i}-1}=0,  \label{m-EOM1}
\\
n+\sum_{b\neq a}\coth\f{u_a-u_b}2 +\sum_i \f{1}{e^{u_a-m_i}-1}=0. \label{u-EOM1}
\ea
One can get useful intuitions by interpreting
these equations as force balance conditions.
The first term in (\ref{m-EOM1}) is a restorative force on $m_i$.
The first term in (\ref{u-EOM1}) represents a constant force applied to
each $u_a$.
These are the external forces acting on the system of eigenvalues.
The other terms are mutual repulsive forces among $m_i$'s and $u_a$'s.

In the absence of $u$-eigenvalues, the $m$-eigenvalues 
obey Wigner's semicircle law
and spread over an interval  of width $2\sqrt\lambda$.
This motivates us to assume that the $m$- and $u$-eigenvalues spread
over regions of length scale $\sqrt\lambda$.
Under this assumption that will be justified a posteriori, 
for large values of $\lambda$, 
we can approximate various expressions as follows:
\ba
\oo{e^{u_a-m_i}-1}&=&
\Bigg\{
\begin{array}{cc}
-1& \hbox{~if~} u_a<m_i \\
0 & \hbox{~if~} m_i<u_a
\end{array},\label{u-m-simplify}
\\
\coth{\f{u_a-u_b}2}
&=&
\Bigg\{
\begin{array}{cc}
-1& \hbox{~if~} u_a<u_b \\
+1 & \hbox{~if~}u_b<u_a
\end{array}.\label{u-u-simplify}
\ea

\begin{figure}[btp]
\begin{center}
\psfrag{N-n}{$N-n$}
\psfrag{l}{$k$}
\psfrag{n}{$n$} 
\includegraphics[width=60mm]{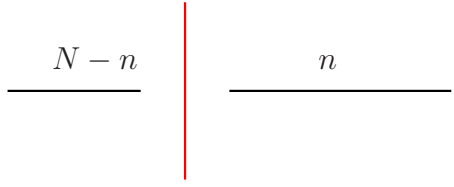} 
\caption{
The eigenvalue distributions in the rectangular case.
The $m$-eigenvalues shown as black lines
split into two groups, $\{m_i^{(1)}\}$
on the right and $\{m_i^{(2)}\}$ on the left.
The $u$-eigenvalues
are distributed uniformly  along the  red line in the
imaginary direction.
}
\label{D5eigenvalues-rec}
\end{center}
\end{figure}

We now propose an eigenvalue distribution
that solves the saddle point equations.
Suppose that the eigenvalues $m_i$ split into
the first group $\{m^{(1)}_i|i=1,...,n\}$ on the right 
and the second one $\{m^{(2)}_i|i=1,...,N-n\}$ on the left.
All the $u_a$'s are between them.
We also assume that the two groups are far enough from each other
so that $1/(m^{(1)}_i-m^{(2)}_j)$ can be ignored.
After we apply (\ref{u-m-simplify}) and (\ref{u-u-simplify}),
(\ref{m-EOM1}) becomes
\ba
-\f{4N}\lambda m^{(1)}_i+2 \sum_{j\neq i}\oo{m^{(1)}_i-m^{(1)}_j}+k=0,
\label{EOM-m(1)}\\
-\f{4N}\lambda m^{(2)}_i+2 \sum_{j\neq i}\oo{m^{(2)}_i-m^{(2)}_j}=0.
\label{EOM-m(2)}
\ea
Then the two groups individually
obey Wigner's semi-circle law.
The first group is spread over an interval of width 
$2\sqrt{g_s n}=\Ocal(\sqrt{\lambda})$
centered at $k\lambda/4N $.
For the second group the interval is centered at the origin 
with width $2\sqrt{g_s(N-n)}=\Ocal(\sqrt\lambda)$.
(\ref{u-EOM1}) simplifies to
\ba
\sum_{b\neq a}\coth\f{u_a-u_b}2=0.
\ea
This equation is solved by $u_a$'s uniformly distributed along a line
in the imaginary direction.
The precise location of the line cannot be determined in this approximation.
See Figure \ref{D5eigenvalues-rec}.

The parallel matrix model problem in the Chern-Simons case
admits an exact solution for the finite 't Hooft parameter \cite{cs-mat}.
It exhibits an essentially identical distribution of eigenvalues.
This gives us confidence in our proposal of the distribution.

We now use the eigenvalue distribution to 
evaluate the Wilson loop vev.
The exponent in (\ref{vev1}) can be evaluated,
with the help of (\ref{u-m-simplify}) and (\ref{u-u-simplify}),
as
\ba
&&
-\f{2N}\lambda \sum_i (m^{(1)}_i)^2+k\sum_i m^{(1)}_i
+\sum_{i<j}\log (m^{(1)}_i-m^{(1)}_j)^2
-\f{2N}\lambda \sum_i ({m^{(2)}_i})^2
\nn\\
&&+\sum_{i<j}\log (m^{(2)}_i-m^{(2)}_j)^2
+\Ocal(N^2\log \lambda)\nn\\
&=&
\f{ n k^2\lambda}{8N}+\Ocal(N^2\log \lambda).
\label{exponent1}
\ea
The last term arises from completion of squares: 
\ba
-(2N/\lambda)( m_i^{(1)})^2
+k m^{(1)}_i=
-(2N/\lambda) (m_i^{(1)}-k\lambda/4N)^2 + k^2\lambda/8N.
\nn
\ea
We thus conclude that
\ba
\langle W_R\rangle_{U(N)}=\exp\left(\f{nk^2\lambda}{8N}+\Ocal (N^2\log\lambda)\right).
\ea

In the $SU(N)$ case, 
we need to work with $M'$ defined in (\ref{Mprime}).
Because of (\ref{TrM'}), the Wilson loop vev is given by
\ba
\langle W_R\rangle_{SU(N)}
=\oo Z \int  dM dU
e^{-\f{2N}{\lambda}\Tr M^2}
\left(\det e^M\right)^{-\f{kn}N}
  \det(1+ e^M\otimes U^\mo)
(\det U)^n.
\ea
(\ref{exponent1}) is replaced by
\ba
-\f{2N}\lambda \sum_i ({m^{(1)}_i})^2+k\sum_i m^{(1)}_i
-\f{kn}N \sum_i m^{(1)}_i
+\sum_{i<j}\log (m^{(1)}_i-m^{(1)}_j)^2
\nn\\
-\f{2N}\lambda \sum_i ({m^{(2)}_i})^2
+\sum_{i<j}\log (m^{(2)}_i-m^{(2)}_j)^2
-\f{kn}N \sum_i m^{(2)}_i.
\label{exponent2}
\ea
The leading parts in $\lambda$ comes from
completion of squares:
\be
-\f{2N}\lambda (m^{(1)}_i)^2+k m^{(1)}_i-\f{kn}N m^{(1)}_i
=-\f{2N}\lambda \left(m^{(1)}_i-\f{\lambda}{4N}k\left(1-\f{n}N\right)\right)^2 
+ \f{\lambda k^2}{8N}\left(1-\f{n}N\right)^2,\nn
\ee
\be
-\f{2N}\lambda (m^{(1)}_i)^2-\f{kn}N m^{(1)}_i
=-\f{2N}\lambda \left(m^{(1)}_i+\f{\lambda}{4N}k\f{n}N\right)^2 
+ \f{\lambda k^2}{8N}\left(\f{n}N\right)^2.\nn
\ee
We find that
\ba
\langle W_R\rangle_{SU(N)}&=& 
\exp \left(
n\f{\lambda k^2}{8N}\left(1-\f{n}N\right)^2
+(N-n)\f{\lambda k^2}{8N}\left(\f{n}N\right)^2
+\Ocal(N^2\log\lambda)
\right)
\nn\\
&=&\exp\left(
\f{\lambda k^2}{8N^2}n(N-n)+\Ocal(N^2\log\lambda)\right).
\ea
This is the special case of (\ref{the-results}),
and is invariant under $R\ra\bar R$ or $n\rightarrow N-n$ as it should be.

Recall that each column in $R$ represents a D5-brane.
Since we have as many $u$-eigenvalues as columns,
we identify the eigenvalues with D5-branes.
The back-reaction of the $m$-eigenvalues to the existence of $u$-eigenvalues
then represents the back-reaction of the geometry to the branes,
i.e., geometric transition.
Without operator insertions, there is a single black region
representing $AdS_5\times S^5$.
The hole in the region, which was identified with D5-branes 
in \cite{Yamaguchi:2006te}, is created by the repulsion
of $m_i$ and $u_a$ in the present formulation.
This is the Wilson loop analog of giant graviton branes
represented by a hole in the fermion droplet \cite{Lin:2004nb}.
In the topological and non-critical string literature
\cite{Aganagic:2003qj,Maldacena:2004sn},
it is well-known that the insertion of a determinant operator 
similar to $\det(1+e^M\otimes U)$ represents non-compact D-branes.
In those cases the eigenvalues of $U$ are the moduli of branes.
Here there is a difference: the eigenvalues are first integrated over 
and are fixed only by saddle point equations.

\subsection{Eigenvalue bound states as D3-branes}
If we make use of the identity \cite{Macdonald,Fulton}
\ba
\sum_R  \Tr_R V\Tr_R e^M=\oo{\det(1-V\otimes e^M)} \label{RR-det}
\ea
instead of (\ref{RRT-det}), we have another expression for $\Tr_R e^M$:
\ba
\Tr_R e^M=\int dV \oo{\det(1-V^\mo \otimes e^M)}\Tr_R V.
\ea
$dV$ is the Haar measure on the unitary group.
More precisely, the contours need to be deformed so
that the eigenvalues of $V$ are
larger than those of $e^M$ in magnitude.
We choose $V$ to be an $n\times n$ unitary matrix,
so that 
\ba
\Tr_R V=(\det V)^k
\ea
for the rectangular Young diagram $R$.
We find that the Wilson loop vev is given by another 2-matrix model
\ba
\langle W_R\rangle_{U(N)}&=& 
\oo Z\int dM dV  \exp\left(-\f{2N}{\lambda}\Tr M^2\right)
 \oo{\det(1-V^\mo\otimes e^M)}
(\det V)^k.
\ea
By diagonalizing the matrices, we get
\ba
\langle W_R\rangle_{U(N)}&\propto
&\int \prod_a dv_a\prod_i dm_i
\exp\Bigg[ -\f{2N}{\lambda}\sum_{i=1}^N
m_i^2+\sum_{i<j}\log(m_i-m_j)^2+k\sum_{a=1}^n v_a\nn\\
&&~~~+ \sum_{a<b}\log\left( 2\sinh\f{v_a-v_b}{2}\right)^2
-\sum_{a,i} \log(1- e^{m_i-v_a} )\Bigg].
\label{vev}
\ea
Let us now analyze this model in the limit $N\ra \infty$
with $\lambda, n/N$, and $k/N$ finite.

The saddle point equations  are
\ba
-\f{4N}\lambda m_i+2\sum_{j\neq i}\oo{m_i-m_j}+\sum_a\f{1}{e^{v_a-m_i}-1}=0,  \label{m-EOM2}
\ea
and
\ba
k+\sum_{b\neq a}\coth\f{v_a-v_b}2 -\sum_i \f{1}{e^{v_a-m_i}-1}=0. \label{v-EOM2}
\ea
Again these equations can be interpreted as
force balance conditions for the eigenvalues.
The difference from the previous subsection
is that the interaction between $m_i$ and $v_a$ is {\it attractive}.

Let us study the eigenvalue distribution for finite but large $\lambda$.
The $v$-eigenvalues as a whole are pulled to the right
by a force of magnitude $kn$.
Unlike in the previous subsection,
we cannot apply (\ref{u-m-simplify}) and (\ref{u-u-simplify})
to the terms involving $m_i$ and $v_a$.
If we did, $v_a$ would only be pulled to the right by $m_i$,
and the force balance would not be achieved.
This suggests that each $v_a$ must be very close to $m_i$.

Each $v$-eigenvalue is pulled to the right by a force
of magnitude $k$.
To balance this, we assume that the $v$-eigenvalue and
an $m$-eigenvalue sit very close to each other.
We also assume, as will be justified a posteriori, 
that this distance is much smaller than
all other length scales in the problem.
There are $n$ $v$-eigenvalues and they are paired up with 
as many $m$-eigenvalues $m^{(1)}_i$ ($i=1,...,n$),
forming $n$  {\it bound states}.
The bound states are
denoted by ($m$-$v$)${}_a$,~$a=1,...,n$.
If we order the eigenvalues so that 
$m^{(1)}_1\lesssim v_1<m^{(1)}_2\lesssim v_2<...<m^{(1)}_n\lesssim v_n$,
$v_a$ is pushed to the right by $v_1,...,v_{a-1}$.\footnote{
Strictly speaking, for $b$ such that $0<a-b\lesssim N/\sqrt\lambda$
(\ref{u-u-simplify}) cannot be applied.
The error from ignoring the effects, however,
is cancelled by the error from ignoring the forces
from $v_b$ with $0<b-a\lesssim N/\sqrt\lambda$.
}
It is also pushed to the left by $v_{a+1},...,v_n$,
but these forces are cancelled by the pull of $m^{(1)}_{a+1},...,m^{(1)}_n$.
Thus (\ref{v-EOM2}) in the approximation (\ref{u-u-simplify}) becomes
\ba
k+a-\f{1}{v_a-m^{(1)}_a}+\Ocal(N/\sqrt\lambda)=0.
\ea
The size of the $a$-th bound state is thus $1/(k+a)$.

Assuming that the group of bound states and the group of 
remaining $m$-eigenvalues $\{m^{(2)}_i|i=1,...,N-n\}$
are far enough from each other,
we can replace
(\ref{m-EOM2}) by
(\ref{EOM-m(1)}) and (\ref{EOM-m(2)}).
Since the $m$-eigenvalues are governed by the same equations
as in the previous subsection, their distribution is the same.
The distribution of the $v$-eigenvalues is identical 
to the distribution of $m^{(1)}_i$ on the macroscopic scale.
This then justifies the assumptions made in the argument.
The computation of the Wilson loop vev is also identical.

Each row in the Young diagram represents a D3-brane \cite{Gomis:2006sb,
Gomis:2006im}.
The rectangular diagram $R$ then corresponds to $n$ D3-branes.
Since we have the same number of bound states,
it is natural to identify each bound state with a D3-brane.
This is the analog of an extra droplet in \cite{Lin:2004nb}
where it was interpreted as dual giant graviton branes
wrapping $S^3$ in $AdS_5$ .

\begin{figure}[tb]
\begin{center}
\psfrag{dist1}{$\sim 1/(k+a)$}
\psfrag{dist2}{$\sim \Ocal(\sqrt\lambda/N)$} 
\psfrag{m}{$m^{(1)}_a$}
\psfrag{v}{$v_a$}
\psfrag{m2}{$m^{(1)}_{a+1}$}
\psfrag{v2}{$v_{a+1}$}
\includegraphics[width=120mm]{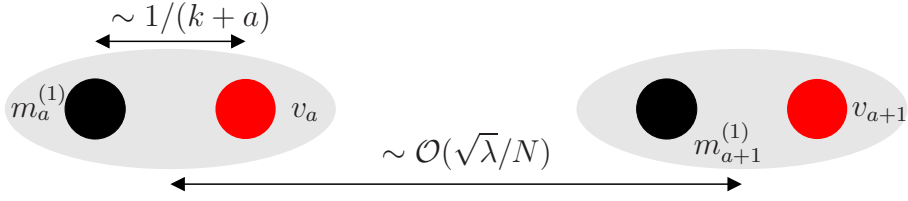}
\caption{
The eigenvalue bound states ($m$-$v$)${}_a$.  
The distance between the two eigenvalues 
in the $a$-th bound state from the left is  $1/(k+a)$.
The distance between neighboring bound states is
much larger and is of the order $\sqrt\lambda/N$.
}
\label{moleculesA}
\end{center}
\end{figure}

\begin{figure}[bt]
\begin{center}
\psfrag{N-n}{$N-n$}
\psfrag{n}{$n$}
\psfrag{m1}{$\{ (m\hbox{-}v)_a\}$}
\psfrag{m2}{$\{m_i^{(2)}\}$}
\includegraphics[width=120mm]{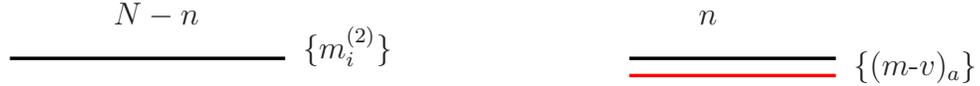}  
\caption{
The eigenvalue distributions for $M={\rm diag}(m_i)_{i=1}^N$ 
and $V={\rm diag}(e^{v_a})_{a=1}^n.$
$\{m_i\}$ split into two groups $\{m_i^{(1)}\}$
and $\{m_i^{(2)}\}$.
The $n$ $v_a$'s are paired with as many $m_i^{(1)}$'s
to form $n$ bound states $(m\hbox{-}v)_a$.
These bound states are distributed according to Wigner's
semicircle law, as represented by the coincident
black and red lines on the right.
The distribution of the remaining $N-n$ $m_i^{(2)}$'s 
independently obeys the semicircle law and is shown on the left.
}
\label{D3-eigenvalues-rec}
\end{center}
\end{figure}

\section{General Young diagram}\label{general}
This section deals with a general Young diagram of the form shown
in Figure \ref{parametrization}, with edge lengths 
$n_I$ and $k_I$ all of order $N$.
We will obtain two multi-matrix models
whose partition functions are the Wilson loop vev.

\subsection{Eigenvalue bound states as D5-branes}
\begin{figure}[htbb]
\begin{center}
\psfrag{R}{$R\equiv R^{(1)}$}
\psfrag{R1}{$R^{(2)}$}
\psfrag{Rm-1}{$R^{(g-1)}$}
\psfrag{Rm}{$R^{(g)}$}
\psfrag{K1}{$K_1$}
\psfrag{K2}{$K_2$}
\psfrag{Km-1}{$K_{g-1}$}
\psfrag{Km}{$K_{g}$}
\includegraphics[width=120mm]{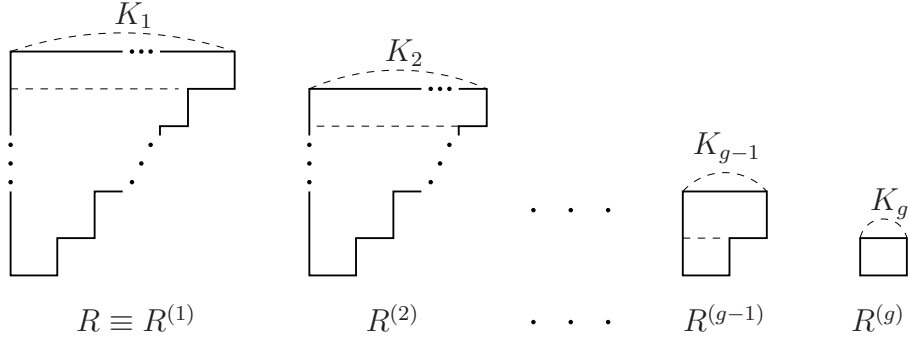}
\caption{
A shrinking sequence of Young diagrams
 $R\equiv R^{(1)}\supset R^{(2)}\supset...\supset R^{(g)}$.
}
\label{D5Young-split}
\end{center}
\end{figure}
Let us first see how to generalize the trick that led to  2-matrix models
in the rectangle case.
We begin with the expression
\ba
\Tr_{R} e^M
=\int dU^{(1)} \det(1+e^M\otimes U^{(1)\mo})\Tr_{R^T} U_1.
\ea
Here $U^{(1)}\in U(K_1)$.
Recall that $K_I\equiv \sum_{J=I}^g k_J$.
(\ref{character}) in this case allows us to write
\ba
\Tr_{R^T} U^{(1)}=(\det U^{(1)})^{n_1}\Tr_{R^{(2)}{}^T} U^{(1)}.
\ea
Here $R^{(2)}$ is obtained by removing the first $n_1$ rows from $R$.
See Figure \ref{D5Young-split}. 
Similarly, 
\ba
\Tr_{R^{(2)}{}^T} U^{(1)}
&=&\int dU^{(2)}
 \oo{\det(1-U^{(1)}\otimes U^{(2)}{}^\mo)}\Tr_{R^{(2)}{}^T} U^{(2)},
\ea
with $U^{(2)}$ in $U(K_2)$.
To be more precise, to avoid singularities 
the integral is performed along
the contours such that the eigenvalues of $U^{(2)}$ are
larger than those of $U^{(1)}$ in magnitude.
This time we have the relation
\ba
\Tr_{R^{(2)}{}^T} U^{(2)}=(\det U^{(2)})^{n_2}\Tr_{R^{(3)}{}^T} U^{(2)}.
\ea
By removing the first $n_2$ rows from $R^{(2)}$
one obtains $R^{(3)}$.
We now repeat the  procedure as many times as we can.
This yields
\ba
&&\Tr_{R} e^M\nn\\
&=&
\hspace{-3mm}
\int
\hspace{-1mm}
 dU^{(1)}
\det(1+e^M\otimes U^{(1)}{}^\mo)(\det U^{(1)})^{n_1}
\hspace{-2mm}
\int
\hspace{-1mm}
 dU^{(2)}
 \oo{\det(1-U^{(1)}\otimes U^{(2)}{}^\mo)}
(\det U^{(2)})^{n_2}\nn\\
&&~~...\int dU^{(g)}
\oo{\det(1-U^{(g-1)}\otimes U^{(g)}{}^\mo)}
(\det U^{(g)})^{n_g}.
\ea
Here $U^{(I)}\in U(K_I)$,
and the integration contours are deformed so that
the eigenvalues of $U^{(I)}$ have larger absolute values
than those of $U^{(I-1)}$.

We thus have
\ba
&&
\left\langle
W_R\right\rangle_{U(N)}
=
\oo Z\int dM  dU^{(1)} dU^{(2)}... dU^{(g)}
e^{-\f{2N}{\lambda}\Tr M^2}
(\det U^{(1)})^{n_1}(\det U^{(2)})^{n_2}...(\det U^{(g)})^{n_g}
\nn\\
&&~~~~~
\times\det(1+e^M\otimes U^{(1)}{}^\mo) 
\oo{\det(1-U^{(1)}\otimes U^{(2)}{}^\mo)}
...
\oo{\det(1-U^{(g-1)}\otimes U^{(g)}{}^\mo)}.
\ea
After redefining $U^{(I)}\ra -U^{(I)}$ and 
diagonalizing the matrices as $M={\rm diag}(m_i)_{i=1}^N$,
$U^{(I)}={\rm diag}(e^{u_a^{(I)}})_{a=1}^{K_I}$, this becomes
\ba
\langle W_R\rangle &\propto&
\int \prod dm_i
\prod du^{(I)}_a
\exp\Bigg[
-\f{2N}{\lambda}\sum_{i=1}^N m_i^2+\sum_{I=1}^g \sum_{a=1}^{K_I} n_I u^{(I)}_a
\nn\\
&&~~+\sum_{i<j}\log(m_i-m_j)^2
+\sum_I \sum_{a<b} \log\left( 2\sinh \f{u^{(I)}_a-u^{(I)}_b}2\right)^2
\nn\\
&&
~~+
\sum_{i,a}\log(1-e^{m_i-u^{(1)}_a}) 
-
\sum_I\sum_{a,b} \log(1-e^{u^{(I-1)}_a-u^{(I)}_b})
\Bigg]. \label{vev-D5-eigen-general}
\ea
Our aim is to understand the behavior in the limit $N\ra \infty$
with $\lambda, n_I/N, k_I/N$ finite, for large values of $\lambda$.
The saddle point equations following from the action are
\ba
&&-\f{4N}{\lambda}m_i+\sum_{j\neq i}\f{2}{m_i-m_j}
-\sum_{a=1}^{K_1} \oo{e^{u^{(1)}_a-m_i}-1}=0 \hbox{~for~}i=1,...,N, 
\\
&&n_1+\sum_{b\neq a}\coth\f{u^{(1)}_a-u^{(1)}_b}2
+\sum_{i=1}^N \oo{e^{u^{(1)}_a-m_i}-1}
+\sum_{b=1}^{K_2} \oo{e^{u^{(2)}_b-u^{(1)}_a}-1}=0 
\nn\\
&&
\hspace{8cm}
\hbox{~for~}a=1,...,K_1,\\
&&
n_{I}+\sum_{b\neq a}\coth\f{u^{(I)}_a-u^{(I)}_b}2
-\sum_{b=1}^{K_{I-1}} \oo{e^{u^{(I)}_a-u^{(I-1)}_b}-1}
+\sum_{b=1}^{K_{I+1}} \oo{e^{u^{(I+1)}_b-u^{(I)}_a}-1}=0
\nn\\
&&
\hspace{6cm}
\hbox{~for~}I=2,...,g-1 \hbox{~and~}a=1,...,K_I,
\ea
and
\ba
&&n_{g}+\sum_{b\neq a}\coth\f{u^{(g)}_a-u^{(g)}_b}2
-\sum_{b=1}^{K_{g-1}} \oo{e^{u^{(g)}_a-u^{(g-1)}_b}-1}
=0
\hbox{~for~} a=1,...,K_g.
\ea
All the terms can be interpreted as forces on eigenvalues.
Note that the interaction between $m_i$ and $u_a^{(1)}$ is repulsive,
while the one between $u_a^{(I)}$ and $u_b^{(I+1)}$ is attractive.
This suggests that some eigenvalues form bound states.

\begin{figure}[htbb]
\begin{center}
\psfrag{u1}{$u^{(1)}$}
\psfrag{u2}{$u^{(2)}$}
\psfrag{ug-1}{$u^{(g-1)}$}
\psfrag{ug}{$u^{(g)}$}
\includegraphics[width=150mm]{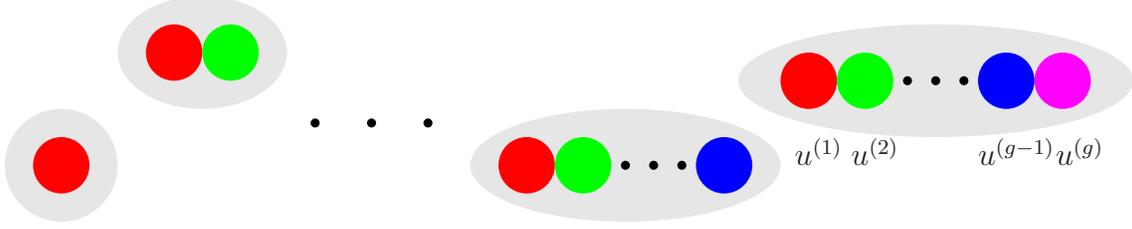}  
\caption{
There are $g$ types of bound states
($u^{(1)}$-$u^{(2)}$-...-$u^{(I)}$),
$I=1,...,g$.
}
\label{moleculesB}
\end{center}
\end{figure}

\begin{figure}[htbb]
\begin{center}
\psfrag{l1}{$n_1$}
\psfrag{l3}{$n_2$}
\psfrag{l2m-3}{$n_{g-1}$} 
\psfrag{l2m-1}{$n_{g}$}
\psfrag{l2m+1}{$n_{g+1}$}
\includegraphics[width=150mm]{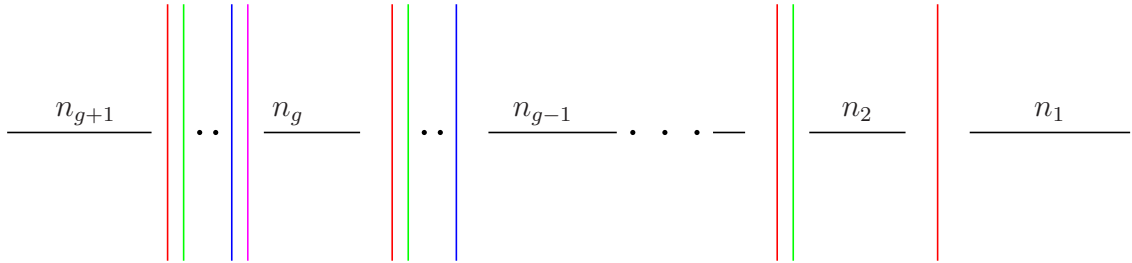} 
\caption{
The eigenvalue distributions for
$M$,
\color{red}$U^{(1)}$\color{black},
\color{green}$U^{(2)}$\color{black},...,
\color{blue}$U^{(g-1)}$\color{black}, and
\color{magenta}$U^{(g)}$\color{black}.
The $N$ $m$-eigenvalues split into $g+1$ groups
$\{m_i^{(I)}|i=1,...,n_I\}$,~$I=1,...,g+1$. 
The $k_I$ ($u^{(1)}$-$u^{(2)}$-...-$u^{(I)}$)-bound states
are distributed uniformly  between $\{m_i^{(I)}\}$
and $\{m_i^{(I+1)}\}$,
and are represented by $I$ coincident colored lines
in the imaginary direction.
}
\label{D5eigenvalues-general}
\end{center}
\end{figure}

Let us work in the $\lambda\ra \infty$ approximation.
The following distribution of eigenvalues solves the
saddle point equations.
It involves $g$ types of bound states.\footnote{
Though it is convenient to talk about bound states,
the size of a bound state is in fact
of the same order $\Ocal(1/N)$ as the distance
to the neighboring bound state.
}
The $I$-th type of bound state, which we denote by
($u^{(1)}$-$u^{(2)}$-...-$u^{(I)}$),
 contains a single $u^{(J)}$-eigenvalue
for $1\leq J\leq I$, as shown in Figure \ref{moleculesB}.
The $m$-eigenvalues split into $g+1$ groups $\{m^{(I)}_i|i=1,..,n_I\}$. 
The $k_I$ ($u^{(1)}$-$u^{(2)}$-...-$u^{(I)}$)-bound states
are distributed along a line in the imaginary direction,
and separate $\{m^{(I)}_i\}$ from $\{m^{(I+1)}_i\}$.
See Figure \ref{D5eigenvalues-general}.
For interactions between two bound states or between a bound state
and an $m$-eigenvalue, we can apply (\ref{u-m-simplify})
and (\ref{u-u-simplify}).
The ($u^{(1)}$-$u^{(2)}$-...-$u^{(I)}$)-bound state
is pulled to the right by an external force of magnitude $n_1+n_2+...+n_I$,
and pushed back to the left by $n_1+n_2+...+n_I$ $m$-eigenvalues
from $\{m^{(J)}_i\}$ for $1\leq J\leq I$.
Bound states of different kinds
do not interact with each other: forces
cancel out among constituent eigenvalues.
The saddle point equations are satisfied if the bound states
are uniformly distributed along  vertical lines.
Calculation of relative positions of constituent eigenvalues 
in a bound state is straightforward.
In the bound state ($u^{(1)}$-$u^{(2)}$-...-$u^{(I)}$),
the distance between the $u^{(J)}$- and $u^{(J+1)}$-eigenvalues
is $1/(\sum_{K=J+1}^I n_K +\sum_{K=J}^{I-1} k_K)=\Ocal(1/N)$
for $1\leq J\leq I-1$.

It seems reasonable to believe, based on the analogy with \cite{cs-mat}, 
that the proposed eigenvalue distribution is the leading saddle point.
Let us calculate the Wilson loop vev by using the proposed distribution.
The exponent of (\ref{vev-D5-eigen-general}) becomes
\ba
&&-\f{2N}{\lambda}\sum_{I=1}^{g+1} \sum_{i=1}^{n_I} (m_i^{(I)})^2
+\sum_{I=1}^g \sum_{i=1}^{n_I} K_I m^{(I)}_i
+\sum_{I=1}^{g+1} \sum_{i<j}\log(m_i^{(I)}-m_j^{(I)})^2
+\Ocal(N^2\log\lambda)
\nn\\
&=&\sum_{I=1}^g \f{\lambda}{8N}n_I K_I^2+\Ocal(N^2\log\lambda).
\label{exponent3}
\ea
We get the final result for the $U(N)$ Yang-Mills:
\ba
\left\langle W_R\right\rangle_{U(N)}
=
\exp\left(
\sum_{I=1}^g \f{\lambda}{8N}n_I K_I^2+\Ocal(N^2\log\lambda)
\right).
\ea

In the $SU(N)$ case, (\ref{TrM'}) allows us to write
\ba
\left\langle
W_R\right\rangle_{SU(N)}
&=&
\oo Z\int dM  dU^{(1)} dU^{(2)}... dU^{(g)}
e^{-\f{2N}{\lambda}\Tr M^2}
(\det e^M)^{-|R|/N}
\nn\\
&&
~~\times
(\det U^{(1)})^{n_1}(\det U^{(2)})^{n_2}...(\det U^{(g)})^{n_g}
\det(1+e^M\otimes U^{(1)}{}^\mo) 
\nn\\
&&
~~\times
\oo{\det(1-U^{(1)}\otimes U^{(2)}{}^\mo)}
...
\oo{\det(1-U^{(g-1)}\otimes U^{(g)}{}^\mo)}.
\ea
(\ref{exponent3}) is replaced by
\ba
&&-\f{2N}{\lambda}\sum_{I=1}^{g+1} \sum_i (m_i^{(I)})^2
+\sum_{I=1}^{g+1} \sum_i (K_I-|R|/N) m^{(I)}_i
\nn\\
&&
~~~
+\sum_{I=1}^{g+1} \sum_{i<j}\log(m_i^{(I)}-m_j^{(I)})^2
+\Ocal(N^2\log\lambda)
\nn\\
&=&\sum_{I=1}^{g+1} \f{\lambda}{8N}n_I (K_I-|R|/N)^2+\Ocal(N^2\log\lambda).
\ea
Recall that $K_{g+1}\equiv 0$.
This is the result (\ref{the-results}) presented in the introduction.

\subsection{Eigenvalue bound states as D3-branes}

\begin{figure}[htbb]
\begin{center}
\psfrag{R}{$R\equiv Q^{(1)}$}
\psfrag{Q2}{$Q^{(2)}$}
\psfrag{Qm-1}{$Q^{(g-1)}$}
\psfrag{Qm}{$Q^{(g)}$}
\psfrag{N1}{$N_1$}
\psfrag{N2}{$N_2$}
\psfrag{Nm-1}{$N_{g-1}$}
\psfrag{Nm}{$N_{g}$}
\includegraphics[width=120mm]{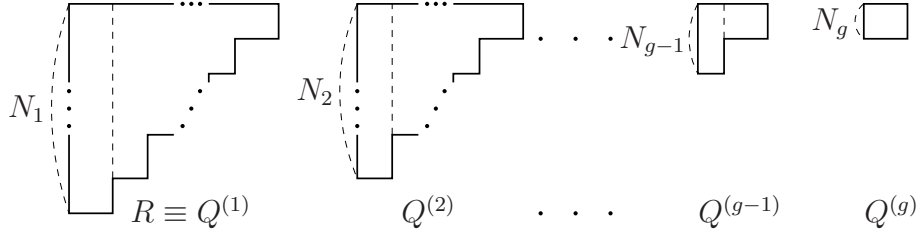}    
\caption{
Another shrinking sequence of Young diagrams 
$R\equiv Q^{(1)}\supset Q^{(2)}\supset...\supset Q^{(g)}$.
}
\label{D3Young-split}
\end{center}
\end{figure}
Another matrix model can be obtained by starting with 
the formula
\ba
\Tr_{R} e^M
=\int dV^{(1)} \oo{\det(1-e^M\otimes V^{(1)}{}^\mo)}\Tr_{R} V^{(1)}.
\ea
Here $V^{(1)}\in U(N_1)$ and $dV^{(1)}$ is the
Haar measure.
We have defined $N_I:=\sum_{J=1}^{g+1-I} n_J$.
Note that 
\ba
\Tr_{R} V^{(1)}=(\det V^{(1)})^{k_g}\Tr_{Q^{(2)}} V^{(1)}.
\ea
Here $Q^{(2)}$ is obtained by removing the first $k_{g}$ rows from $R$.
See Figure \ref{D3Young-split}. 
Similarly, 
\ba
\Tr_{Q^{(2)}} V^{(1)}
&=&\int dV^{(2)}
 \oo{\det(1-V^{(1)}\otimes V^{(2)}{}^\mo)}\Tr_{Q^{(2)}} V^{(2)},
\ea
with $V^{(2)}$ essentially in $U(N_2)$.
To avoid singularities 
the integral is performed along
appropriate contours.
We repeat the same procedure as many times as possible.
This gives
\ba
\Tr_{R} e^M&=&\int dV^{(1)}
\oo{\det(1-e^M\otimes V^{(1)}{}^\mo)}(\det V^{(1)})^{k_{g}}
\int dV^{(2)} \oo{\det(1-V^{(1)}\otimes V^{(2)}{}^\mo)}
\nn\\
&&~
\times
(\det V^{(2)})^{k_{g-1}}...\int dV^{(g)}
\oo{\det(1-V^{(g-1)}\otimes V^{(g)}{}^\mo)}
(\det V^{(g)})^{k_{1}}.
\ea
Here $V^{(I)}\in U(N_I)$,
and the integration contours are deformed so that
the eigenvalues of $V^{(I)}$ have larger absolute values
than $V^{(I-1)}$.

We thus have
\ba
\langle W_R\rangle
\hspace{-1mm}
=
\hspace{-1mm}
\oo Z
\hspace{-1mm}
\int
\hspace{-1mm}
 dM  dV^{(1)} dV^{(2)}... dV^{(g)}
e^{-\f{2N}{\lambda}\Tr M^2}
(\det V^{(1)})^{k_{g}}(\det V^{(2)})^{k_{g-1}}
...(\det V^{(g)})^{k_{1}}
\nn\\
\times \oo{\det(1-e^M\otimes V^{(1)}{}^\mo)}
\oo{\det(1-V^{(1)}\otimes V^{(2)}{}^\mo)}
...
\oo{\det(1-V^{(g-1)}\otimes V^{(g)}{}^\mo)}.
\ea
In terms of the eigenvalues, the Wilson loop is given by
\ba
\langle W_R\rangle&\propto&
\int \prod dm_i
\prod dv^{(I)}_a
\exp\Bigg[
-\f{2N}{\lambda}\sum_{i=1}^N m_i^2+\sum_{I=1}^g \sum_{a=1}^{N_I} 
k_{g+1-I}v^{(I)}_a
\nn\\
&&~~+\sum_{i<j}\log(m_i-m_j)^2
+\sum_I \sum_{a<b} \log\left( 2\sinh \f{v^{(I)}_a-v^{(I)}_b}2\right)^2
\nn\\
&&~~-\sum_{i,a}\log(1-e^{m_i-v^{(1)}_a}) 
-\sum_I\sum_{a,b} \log(1-e^{v^{(I-1)}_a-v^{(I)}_b})
\Bigg].
\ea
Let us study  this model in the large $N$ limit
with $n_I/N$ and $k_I/N$ fixed.
The saddle point equations are 
\ba
&&-\f{4N}{\lambda}m_i+\sum_{j\neq i}\f{2}{m_i-m_j}
+\sum_{a=1}^{N_1} \oo{e^{v^{(1)}_a-m_i}-1}=0 \hbox{~for~}i=1,...,N,\\
&&k_{g}+\sum_{b\neq a}\coth\f{v^{(1)}_a-v^{(1)}_b}2
-\sum_{i=1}^N \oo{e^{v^{(1)}_a-m_i}-1}
+\sum_{b=1}^{N_2} \oo{e^{v^{(2)}_b-v^{(1)}_a}-1}=0 
\nn\\
&&
\hspace{8cm}
\hbox{~for~}
a=1,..., N_1,
\\
&&k_{g+1-I}+\sum_{b\neq a}\coth\f{v^{(I)}_a-v^{(I)}_b}2
-\sum_{b=1}^{N_{I-1}} \oo{e^{v^{(I)}_a-v^{(I-1)}_b}-1}
+\sum_{b=1}^{N_{I+1}} \oo{e^{v^{(I+1)}_b-v^{(I)}_a}-1}=0
\nn\\
&&
\hspace{6cm}
\hbox{~for~}I=2,...,g-1 \hbox{~and~}a=1,...,N_I,
\ea
and
\ba
&&k_{1}+\sum_{b\neq a}\coth\f{v^{(g)}_a-v^{(g)}_b}2
-\sum_{b=1}^{N_{g-1}} \oo{e^{v^{(g)}_a-v^{(g-1)}_b}-1}
=0
\hbox{~for~} a=1,...,N_g,
\ea 
The forces between $m_i$ and $v^{(1)}_a$
as well as between $v^{(I)}_a$ and $v^{(I+1)}_b$
are attractive.

\begin{figure}[htbb]
\begin{center}
\psfrag{m}{$m$}
\psfrag{v1}{$v^{(1)}$}
\psfrag{v2}{$v^{(2)}$}
\psfrag{vI}{$v^{(g-1)}$}
\includegraphics[width=150mm]{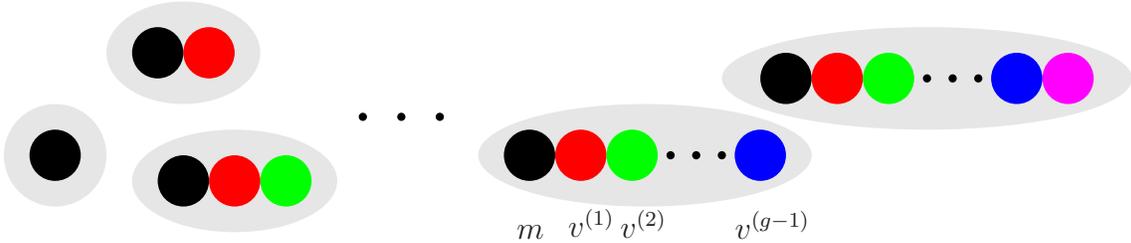}  
\caption{
There are $g+1$ types of bound states  
($m$-$v^{(1)}$-...-$v^{(I)}$),
$I=0,1,...,g$.
}   
\label{moleculesC}
\end{center}
\end{figure}

Our proposed distribution of the eigenvalues
for large $\lambda$
is as follows.
There are $g$ types of bound states denoted by
($m$-$v^{(1)}$-...-$v^{(I)}$) for $I=0,1,...,g$.
Each bound state of the $I$-th type
has a single $m$-eigenvalue
as well as a single $v^{(J)}$-eigenvalue for  $1\leq J\leq I$.
The size of a bound state
is of the order $1/N$, and is much smaller
than the distance between two neighboring bound states.
The $n_{g+1-I}$ ($m$-$v^{(1)}$-...-$v^{(I)}$)-bound states
are distributed according to the semi-circle law
centered at $\lambda K_I /4N$.
Using (\ref{u-m-simplify}) and (\ref{u-u-simplify}),
one can confirm that the force balance among bound states
is achieved.
It is straightforward to calculate the
distance between eigenvalues in a bound state.
In the $a$-th ($m$-$v^{(1)}$-...-$v^{(I)}$)
bound state,
the distance between the $v^{(J)}$- and $v^{(J+1)}$-eigenvalues turns out to be
$1/((I-J)a+\sum_{K=J}^I k_{g+1-K})=\Ocal(1/N)$ for $0\leq J\leq I-1$.
Here $v^{(0)}\equiv m$  and $k_{g+1}\equiv 0$.
This configuration solves the saddle point equations.
The distribution of the $m$-eigenvalues is identical
to the one in the previous subsection,
and leads to the same results for the Wilson loop vev.

\begin{figure}[htbb]
\begin{center}
\psfrag{l1}{$n_1$}
\psfrag{l3}{$n_2$}
\psfrag{l2m-3}{$n_{g-1}$} 
\psfrag{l2m-1}{$n_{g}$}
\psfrag{l2m+1}{$n_{g+1}$}
\includegraphics[width=120mm]{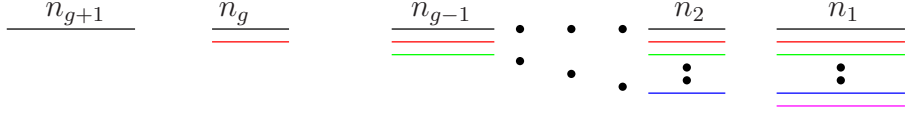}  
\caption{
The eigenvalue distributions for
$M$,
\color{red}$V^{(1)}$\color{black},
\color{green}$V^{(2)}$\color{black},...,
\color{blue}$V^{(g-1)}$\color{black},
\color{magenta}$V^{(g)}$\color{black}.
The $n_{g+1-I}$ 
($m$-$v^{(1)}$-...-$v^{(I)}$)-bound states
are distributed according to Wigner's semi-circle law.
}
\label{D3eigenvalues}
\end{center}
\end{figure}

\section*{Acknowledgments}
I thank Jaume Gomis and Kentaroh Yoshida for helpful conversations.
I am especially grateful to 
Sean Hartnoll for many useful discussions, and 
Nick Halmagyi for collaboration on a
related project \cite{cs-mat}.
I also acknowledge the hospitality of the Simons Workshop at Stony Brook,
where this work was completed.
My research is supported in part by the NSF grants PHY-05-51164 and PHY-04-56556.





\bibliography{circ10}

\end{document}